\documentclass[aps,showpacs,11pt,nofootinbib]{revtex4}
\usepackage{dcolumn}
\usepackage{graphicx}
\usepackage{amsmath}
\usepackage{amsfonts}
\usepackage{amssymb}
\usepackage{psfrag}
\usepackage{wrapfig}
\usepackage{subfigure}
\usepackage{makeidx}
\usepackage{bm}
\usepackage{epsf}
\usepackage{hyperref}
\usepackage{color}
\usepackage{multirow}

\begin{document}

\title{Tachyonic Instability of Reissner-Nordstr\"{o}m-Melvin Black Holes in Einstein-Maxwell-Scalar Theory}

\author{Hengyu Xu$^1$}
\email{xuhengyu0501@outlook.com}
\author{Shao-Jun Zhang$^{1,2}$}
\email{sjzhang@zjut.edu.cn}
\affiliation{$^1$ College of Science, Zhejiang University of Technology, Hangzhou 310032, China\\
	$^2$Institute for Theoretical Physics $\&$ Cosmology, Zhejiang University of Technology, Hangzhou 310032, China}
\date{\today}

\begin{abstract}
	\indent In the framework of Einstein-Maxwell-scalar theory, we studied scalar field perturbations of Reissner-Nordstr\"{o}m-Melvin (RNM) black holes, which describes the RN black holes immersed in a uniform magnetic field. Due to the coupling to the Maxwell term, the scalar field acquires an effective mass whose square, in the presence of the magnetic field, will become negative somewhere outside the horizon for either sign of the coupling constant $\alpha$, thus triggering the tachyonic instability and leading to spontaneous scalarization when $\alpha$ is large enough. The magnetic field has significant influences on the waveforms and the onset of the instability, which differs for different sign of $\alpha$. Effects of the black hole charge on the instability are also studied.
\end{abstract}


\maketitle

\section{Introduction}

Recently, phenomenon of spontaneous scalarization over black holes (BHs) has been attracting lots of attention. It provides a novel mechanism to endow BHs with additional scalar hair, thus circumventing the famous BH no-hair theorem \cite{Bekenstein:1998aw,Robinson:2004zz,Chrusciel:2012jk} while keeping consistent with the current available astrophysical observations. Actually, this phenomenon was first observed long time ago by Damour and Esposito-Farese in neutron stars in the framework of a scalar-tensor theory \cite{Damour:1993hw}. In this theory, stars with vanishing scalar field are solutions, over which the scalar field perturbation acquires a negative effective mass square due to its coupling to the surrounding matter, thus triggering the tachyonic instability and leading to the spontaneous scalarization over the stars. This phenomenon is then realized in the scenario of BHs. In Refs. \cite{Doneva:2017bvd,Silva:2017uqg,Antoniou:2017acq,Cunha:2019dwb}, the authors considered the Einstein-scalar-Gauss-Bonnet (EsGB) theory where the scalar field is coupled to the Gauss-Bonnet (GB) invariant non-minimally through some appropriate coupling function. As general relativity (GR), EsGB also allows the standard Kerr BH with no scalar hair. However, differently, the Kerr metric is not the only one describing a rotating BH in EsGB. Rather, there exists another branch of BH solutions carrying scalar hair in this theory. The bald Kerr BH can be spontaneously scalarized and transformed into the hairy BH branch. In such process, due to the coupling between the scalar field and the GB invariant, the effective mass square of the scalar field perturbations becomes negative in the strong curvature regime, thus triggers the tachyonic instability, and addresses the Kerr BH with scalar hair.

This phenomenon is also found to exist in other modified gravity theories, such as the dynamical Chern-Simon gravity (dCSG) in which the scalar field is coupled instead to the gravitational Chern-Simons invariant \cite{Gao:2018acg,Myung:2020etf,Doneva:2021dcc,Zhang:2021btn,Chatzifotis:2022mob}. Inspired by the work of Damour and Esposito-Farese mentioned above, spontaneous scalarization over BH backgrounds induced by the surrounding matter fields instead of the curvature is also realized in Einstein-Maxwell-scalar (EMs) theory \cite{Herdeiro:2018wub}, where the standard bald Reissner-Nordstr\"{o}m (RN) BH can be spontaneously scalarized due to the non-minimal coupling between the scalar field and the Maxwell term. Inspired by these work, amounts of effort have been devoted to this direction by considering variants of scalarization models, various forms of coupling functions, properties and astrophysical implications of the scalarized BHs, the so-called spin-induced spontaneous scalarization, spontaneous scalarization in binary systems, etc. For more details, see a recent review \cite{Doneva:2022ewd} and references therein. The possible existences of scalarized BHs challenge the Kerr hypothesis of GR, which claims that the most general BH should be characterized only by the mass, angular momentum and the charge. With the great advances of astrophysical observations in recent years, especially in gravitational wave (GW) detection \cite{LIGOScientific:2016aoc,LIGOScientific:2016sjg,LIGOScientific:2017bnn,LIGOScientific:2017ycc,LIGOScientific:2020iuh,LIGOScientific:2021usb} and at a variety of electromagnetic wavelengths observations \cite{GRAVITY:2020gka,EventHorizonTelescope:2019dse,EventHorizonTelescope:2021srq}, one can expect to examine this phenomenon and thus test GR in the near future.

However, most of the studies on spontaneous scalarization to date have considered rather ideal situations in which BHs live in a clean vacuum. In the realistic universe, BHs always live in rather complicated astrophysical environments, such as the magnetic fields, accretion disks, dark matter halos, cosmological expansion, etc. These environments can influence physical processes around BHs considerably \cite{Blandford:1977ds,Barausse:2014tra,Brito:2014nja}. Then, it is natural and important to consider the impact of the realistic astrophysical environments on the BH scalarization. Among these various environmental factors, our interest in this work is in the magnetic fields. According to available astrophysical observations, magnetic field is believed to pervade our universe with different amplitudes on different scales \cite{Crocker:2010xc,Olausen:2013bpa}. Moreover, astrophysical observations of recent years also support the existences of BHs permeating in the magnetic backgrounds. For example, it has been observed that the supermassive BH Sagittarius $A^\ast$ in our galaxy is accompanied by the magnetar SGR J$1745$-$2900$ \cite{Mori:2013yda,Kennea:2013dfa,Eatough:2013nva,Olausen:2013bpa}, and there is a strong magnetic field around $M 87^\ast$ \cite{EventHorizonTelescope:2021srq}.

There are already some preliminary studies considering the influences of magnetic fields on spontaneous scalarization. In Refs. \cite{Annulli:2022ivr,Hod:2022qir}, in the framework of EsGB, the authors considered the possible existence of tachyonic instability and spontaneous scalarization over the Kerr-Newmann-Melvin (KNM) BHs. This kind of BHs describes a Kerr-Newmann BH immersed in a uniform magnetic field \cite{Wald:1974np,Ernst:1976mzr,Ernst-Wild:1976,Gibbons:2013yq}, and can be taken as a simple model to draw a qualitative picture of the interactions between BHs and the magnetic field surrounding it. They found that the presence of the magnetic field will push up the spin threshold for the tachyonic instability. The work is then extended to dCSG theory in Ref. \cite{Zhang:2022sgt}. By performing time evolutions of the scalar field perturbations over the Reissner-Nordstr\"{o}m-Melvin (RNM) BHs, the authors found that the tachyonic instability and spontaneous scalarization can be induced by the magnetic field. Moreover, the magnetic field will change the waveforms and the ringdown modes dramatically. Interestingly, it was found in Ref. \cite{Brihaye:2021jop} that even in the absence of compact objects (like BHs), scalar field can still condensate in the Melvin magnetic universe when it is coupled non-minimally either to the curvature or to the magnetic field.

In this work, we would like to extend the above studies to EMs theory where the spontaneous scalarization over BHs will be triggered by the Maxwell field instead of the curvature. We assume the RNM BHs as the background to study the influences of magnetic fields. We will show that for different spontaneous scalarization models, influences of magnetic fields will be rather different. The work is organized as follows. In Sec. II, we give a brief introduction of the EMs theory and the RNM BH, and derive the scalar field perturbation equations. In Sec. III, we describe the numerical strategy to solve the scalar field perturbation equation and report our numerical results in detail for different sign of coupling constant. The final section is devoted to the Summary and Discussion.

In this work, we use the units $c=G =4\pi \epsilon_0 =1$, where $c, G, \epsilon_0$ are the the speed of light in vacuum, the Newton gravitational constant and the vacuum permittivity respectively.

\section{The model}

The general action of Einstein-Maxwell-scalar (EMs) theory is
\begin{eqnarray}
	S&=& -\int dx^4\sqrt{-g}\left[R - h(\phi)F^2-2\partial_\mu\phi\partial^\mu\phi\right],
\end{eqnarray}
where the scalar field is non-minimally coupled to the Maxwell term $F^2 \equiv F_{\mu\nu}F^{\mu\nu}$ through the coupling function $h(\phi)$. By varying the action, one can derive the equations of motion
\begin{eqnarray}
	&&\nabla_\mu \nabla^\mu\phi =\frac{1}{4}\frac{d h}{d\phi}F^2,\label{ScalarEq}\\
	&&\nabla (h F^{ab}) = 0, \label{GaugeEq}\\
	&&R_{\mu\nu}-\frac{1}{2}g_{\mu\nu}R = h T^{EM}_{\mu\nu} + T^\phi_{\mu\nu},\label{MetricEq}\\
	&&T^{EM}_{\mu\nu}= 2 F_{\mu \rho} F_\nu^{~\rho} - \frac{1}{2} g_{\mu \nu} F^2,\nonumber\\
	&&T^\phi_{\mu\nu} = 2 \nabla_\mu \phi \nabla_\nu \phi - g_{\mu\nu} \nabla_\rho \phi \nabla^\rho \phi. \nonumber
\end{eqnarray}
As Ref. \cite{Herdeiro:2018wub}, we take the form of the coupling function as
\begin{eqnarray}
	h(\phi) = e^{-\alpha \phi^2},
\end{eqnarray}
where $\alpha$ is a coupling constant. With this choice, the theory admits the Reissner-Nordstr\"{o}m-Melvin (RNM) solution with vanishing scalar field $\phi=0$ \cite{Ernst:1976mzr,Gibbons:2013yq}
\begin{eqnarray}
	ds^2 &=& H \left[-f dt^2 + f^{-1} dr^2 + r^2 d\theta^2\right] + H^{-1} r^2 \sin^2\theta(d\varphi - \Omega dt)^2, \label{RNMelvin}\\
	A_\mu dx^\mu &=& \Phi_0 dt + \Phi_3 (d\varphi - \Omega dt),
\end{eqnarray}
where
\begin{eqnarray}
	f &=& 1 - \frac{2 M}{r} + \frac{Q^2}{r^2},\nonumber\\
	H &=& 1 + \frac{1}{2} B^2 \left(r^2 \sin^2\theta + 3 Q^2 \cos^2\theta\right) + \frac{1}{16} B^4 \left(r^2 \sin^2\theta + Q^2 \cos^2\theta\right)^2,\nonumber\\
	\Omega &=& -\frac{2 Q B}{r} + \frac{Q B^3 r}{2} \left(1 + f \cos^2\theta\right),\nonumber\\
	\Phi_0 &=& -\frac{Q}{r} + \frac{3}{4} Q B^2 r \left(1 + f \cos^2\theta\right),\nonumber\\
	\Phi_3 &=& \frac{2}{B} - H^{-1} \left[\frac{2}{B} + \frac{B}{2} \left(r^2 \sin^2\theta + 3 Q^2 \cos^2\theta\right)\right].
\end{eqnarray}
This GR electro-vacuum solution describes a RN black hole with charge $Q$ immersed in a uniform magnetic field $B$ aligned along the symmetry axis. Although this BH solution is not asymptotically flat but resembles a magnetic Melvin-like universe \cite{Gibbons:2013yq,Melvin:1963qx}, we can take it as a simple model to draw a qualitative picture of the interactions between the BH and the magnetic field surrounding it. The event horizon is located at $r=r_+ = M + \sqrt{M^2 - Q^2}$. On this background, we would like to study the scalar field perturbation whose dynamics is governed by Eq. (\ref{ScalarEq}) evaluated on the background, that is
\begin{eqnarray}
	\nabla^2\phi=m^2_{\rm eff} \phi, \label{ScalarEq1}
\end{eqnarray}
where the scalar field acquires an effective mass whose square is $m^2_{\rm eff}=- \frac{1}{2} \alpha F^2$ with the Maxwell term as
\begin{equation}
	F^2 = - \frac{2 (\partial_\theta \Phi_0 - \Phi_3 \partial_\theta \Omega)^2}{f H^2 r^2} + \frac{2 (\partial_\theta \Phi_3)^2}{r^4 \sin^2 \theta} - \frac{2 (\partial_r \Phi_0 - \Phi_3 \partial_r \Omega)^2}{H^2} + \frac{2 f (\partial_r \Phi_3)^2}{r^2 \sin^2\theta}. \label{MaxwellTerm}
\end{equation}
Note that $m^2_{\rm eff}$ is position-dependent and depends on parameters $(M, Q, B)$. Following are several properties of the Maxwell term (\ref{MaxwellTerm}) which are useful for our discussions below:
\begin{itemize}
	\item It is even in the $\theta$-direction with respect to $\theta = \frac{\pi}{2}$
	      \begin{equation}
		      \theta \rightarrow \pi - \theta \quad \Rightarrow \quad F^2 \rightarrow F^2.
	      \end{equation}
	\item The full expression of $F^2$ is rather involved. We can get a first sense of the influence of the magnetic field on $F^2$ from the small-$B$ expansion,
	      \begin{equation}
		      F^2 = -\frac{2 Q^2}{r^4} + \left(2 + \frac{39 Q^4 \cos^2 \theta}{r^4} - \frac{Q^2 (13 + 23 \cos 2\theta) + 8 M r \sin^2\theta}{2 r^2}\right) B^2 + {\cal O}(B^4).
	      \end{equation}
	\item In the limit $B=0$, the RNM BH reduces to the RN BH. And the Maxwell term reduces to
	      \begin{equation}
		      F^2 = - \frac{2 Q^2}{r^4},\qquad B \rightarrow 0,
	      \end{equation}
	      which is always negative for $Q \neq 0$. In this case, only for $\alpha < 0$ can $m^2_{\rm eff}$ become negative and the tachyonic instability be triggered. This case has been discussed thoroughly in past few years \cite{Herdeiro:2018wub}.

	      While in the limit $Q = 0$, the RNM BH reduces to the Schwarzschild-Melvin BH \cite{Ernst:1976mzr} and the Maxwell term becomes
	      \begin{equation}
		      F^2 = \frac{512 B^2 (r -2 M \sin^2 \theta)}{r (B^2 r^2 \sin^2 \theta +4)^4}, \qquad Q \rightarrow 0
	      \end{equation}
	      which is always positive for $B \neq 0$. In this case, only for $\alpha > 0$ can $m^2_{\rm eff}$ become negative and the tachyonic instability be triggered.

	      \begin{figure}[!htbp]
		      \centering
		      \includegraphics[width=0.47\textwidth]{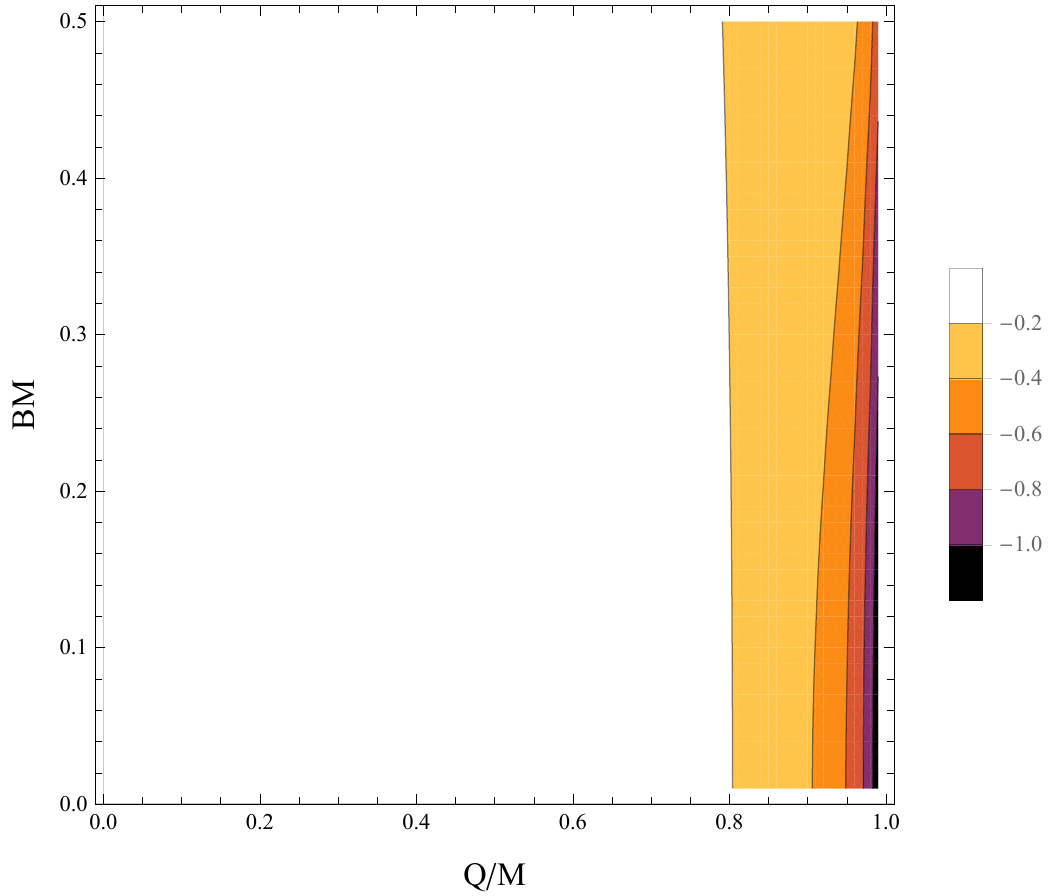}\quad
		      \includegraphics[width=0.47\textwidth]{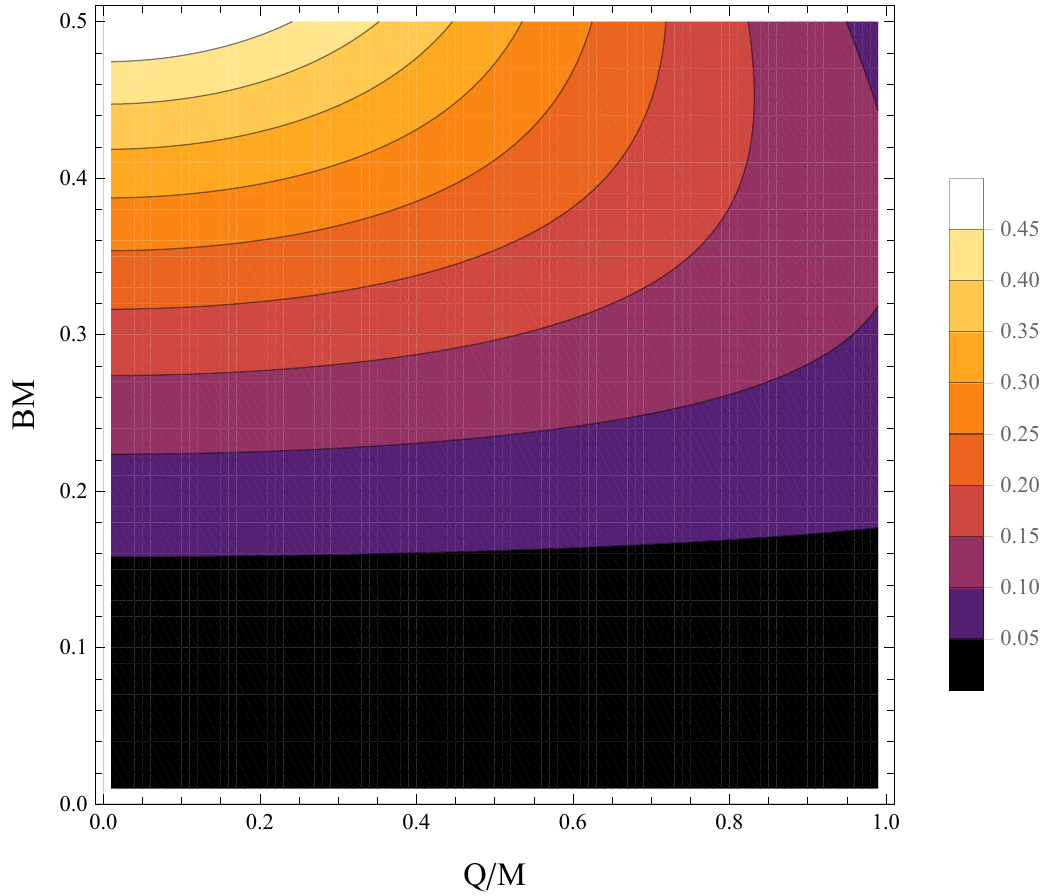}
		      \caption{(color online) Minimum value (left panel) and maximum value (right panel) of the Maxwell term $F^2$ outside the horizon in the $\{Q/M, BM\}$-phase space.}
		      \label{fig:F2MinMax}
	      \end{figure}

	\item For the general case $B \neq 0$ and $Q \neq 0$, $F^2$ is positive somewhere while negative elsewhere. In Fig. \ref{fig:F2MinMax}, we plot the extremal values of the Maxwell term $F^2$ outside the horizon in the $\{Q/M, BM\}$-phase space. From the figure, one can see that the minimum value of $F^2$ is always negative while the maximum value of $F^2$ is always positive. This means that, when the magnetic field is present, for either sign of $\alpha$ can $m^2_{\rm eff}$ become negative somewhere thus triggering the tachyonic instability when the coupling constant $\alpha$ is large enough. This is different from the $B=0$ case and $Q=0$ case.

\end{itemize}

In the following section, we will study carefully the time evolution of the scalar field perturbation by solving Eq. (\ref{ScalarEq1}) and obtain object pictures of the influences of the magnetic field $B$, the coupling constant $\alpha$ and also the BH charge $Q$ on wave dynamics.

\section{Numerical strategy and results}

We will apply the numerical method as Refs. \cite{Krivan:1996da,PazosAvalos:2004rp,Dolan:2011dx,Doneva:2020nbb,Doneva:2020kfv,Schiesser,Zhang:2022sgt} to solve the scalar field perturbation equation (\ref{ScalarEq1}). As Ref. \cite{Zhang:2022sgt}, we introduce the tortoise coordinate $x$ to map the radial domain $r \in (r_+, +\infty)$ to $x \in (-\infty, +\infty)$, and a Kerr-like azimuthal coordinate $\tilde{\varphi}$ to remove unphysical pathology induced by the coordinate $\varphi$ near the horizon \cite{Krivan:1996da}. The two new coordinates are defined respectively as
\begin{equation}
	dx \equiv dr/f,\quad d \tilde{\varphi} = d \varphi + \Omega dx.
\end{equation}
Then the scalar field perturbation equation (\ref{ScalarEq1}) evaluated on the background becomes
\begin{eqnarray}
	&&r^2 \left(\partial_t^2 - \partial_x^2\right) \phi- 2 f r \partial_x \phi+ 2 r^2 \Omega \left(\partial_t \partial_{\tilde{\varphi}} - \partial_x \partial_{\tilde{\varphi}}\right) \phi \nonumber\\
	&&- \partial_x (r^2 \Omega) \partial_{\tilde{\varphi}} \phi - \frac{f}{\sin\theta} \partial_\theta \left(\sin\theta \partial_\theta \phi\right) -\frac{f H^2}{\sin^2 \theta} \partial_{\tilde{\varphi}}^2 \phi  = - m_{\rm eff}^2 \ f H r^2 \phi.
\end{eqnarray}
By decomposing the scalar field perturbation, $\phi (t, x, \theta, \tilde{\varphi}) = \sum_m \Psi (t, x, \theta) e^{i m \tilde{\varphi}}$, and introducing an auxiliary variable $\Pi \equiv \partial_t \Psi$, the perturbation equation can be cast eventually into a form of two coupled first-order partial differential equations
\begin{eqnarray}
	\partial_t \Psi = &&\Pi,\nonumber\\
	\partial_t \Pi = &&- 2 i m \Omega \Pi + \partial_x^2 \Psi + 2 \left(i m \Omega + \frac{f}{r}\right) \partial_x \Psi + \frac{f}{r^2} \left(\partial_\theta^2 + \cot\theta \partial_\theta\right)\Psi \nonumber\\
	&&+\left(\frac{i m}{r^2} \partial_x (r^2 \Omega) - \frac{m^2 f H^2}{r^2 \sin^2\theta}  - m_{\rm eff}^2 \ f H\right) \Psi,\label{ScalarEqFinal}
\end{eqnarray}
which can be solved by the method of line \cite{Schiesser}. To solve the perturbation equation, physical boundary conditions are needed. We impose ingoing wave condition at the horizon following Ref. \cite{Ruoff2000}, while impose Dirichlet boundary condition at a large radial cutoff with order of $x \sim 1/B$ \cite{Brito:2014nja}. Moreover, at the poles $\theta=0$ and $\pi$, physical boundary conditions $\Psi|_{\theta=0, \pi} = 0$ for $m \neq 0$ while $\partial_\theta \Psi|_{\theta=0, \pi} = 0$ for $m=0$ \cite{Dolan:2011dx} are imposed. For more details, please refer to Refs. \cite{Krivan:1996da,PazosAvalos:2004rp,Dolan:2011dx,Doneva:2020nbb,Doneva:2020kfv,Schiesser,Zhang:2022sgt}.

As Refs. \cite{Zhang:2020pko,Zhang:2021btn,Zhang:2022sgt}, taking into account the ``mode-mixing phenomenon" \cite{Thuestad:2017ngu,Zenginoglu:2012us,Burko:2013bra} and for simplicity, we consider $m=0$ and assume the initial scalar field perturbation as
\begin{eqnarray}
	\Psi (t=0, x) &\sim& e^{-\frac{(x - x_c)^2}{2 \sigma^2}},\\
	\Pi (t=0, x) &=& 0,
\end{eqnarray}
which is a Gaussian wave-packet localized outside the horizon at $x=x_c$ with width $\sigma$ and has time symmetry. Also, we set $M = 1$ so that all quantities are measured in units of $M$. Without loss of generality, observers are assumed to locate at $x = 10 M$ and $\theta = \frac{\pi}{5}$.

\subsection{$\alpha<0$}

\begin{figure}[!htbp]
	\includegraphics[width=0.45\textwidth]{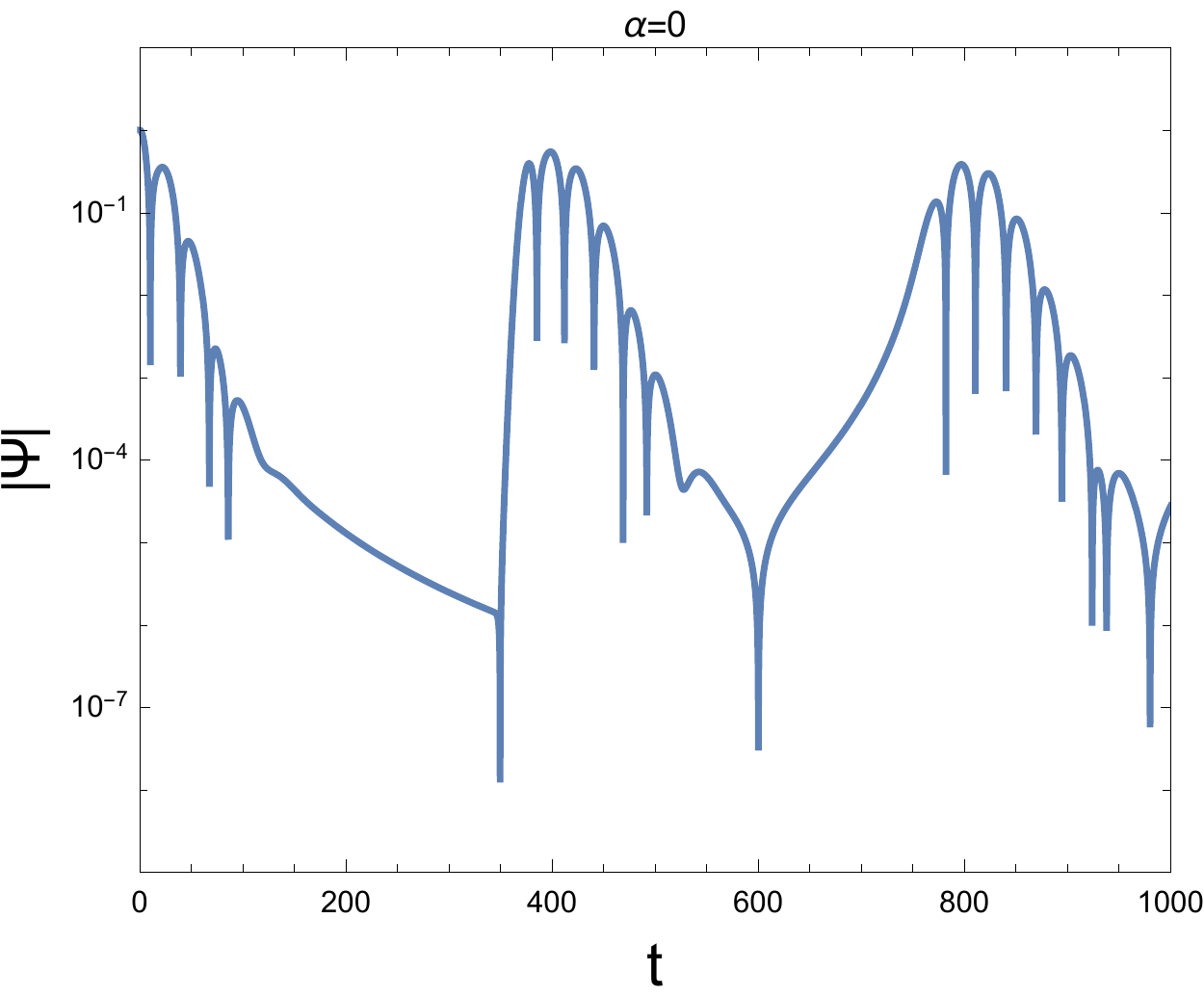}\quad
	\includegraphics[width=0.45\textwidth]{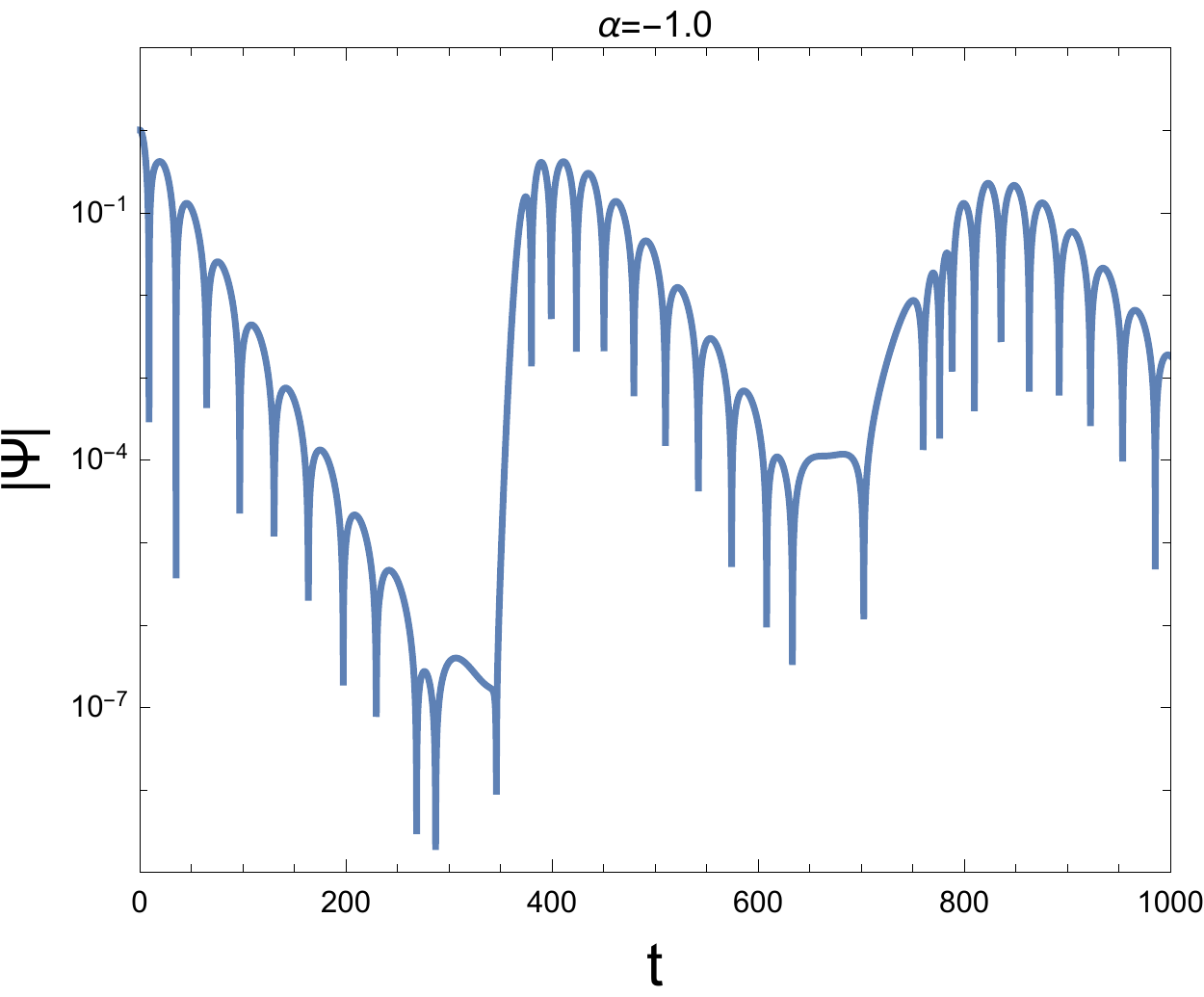}
	\includegraphics[width=0.45\textwidth]{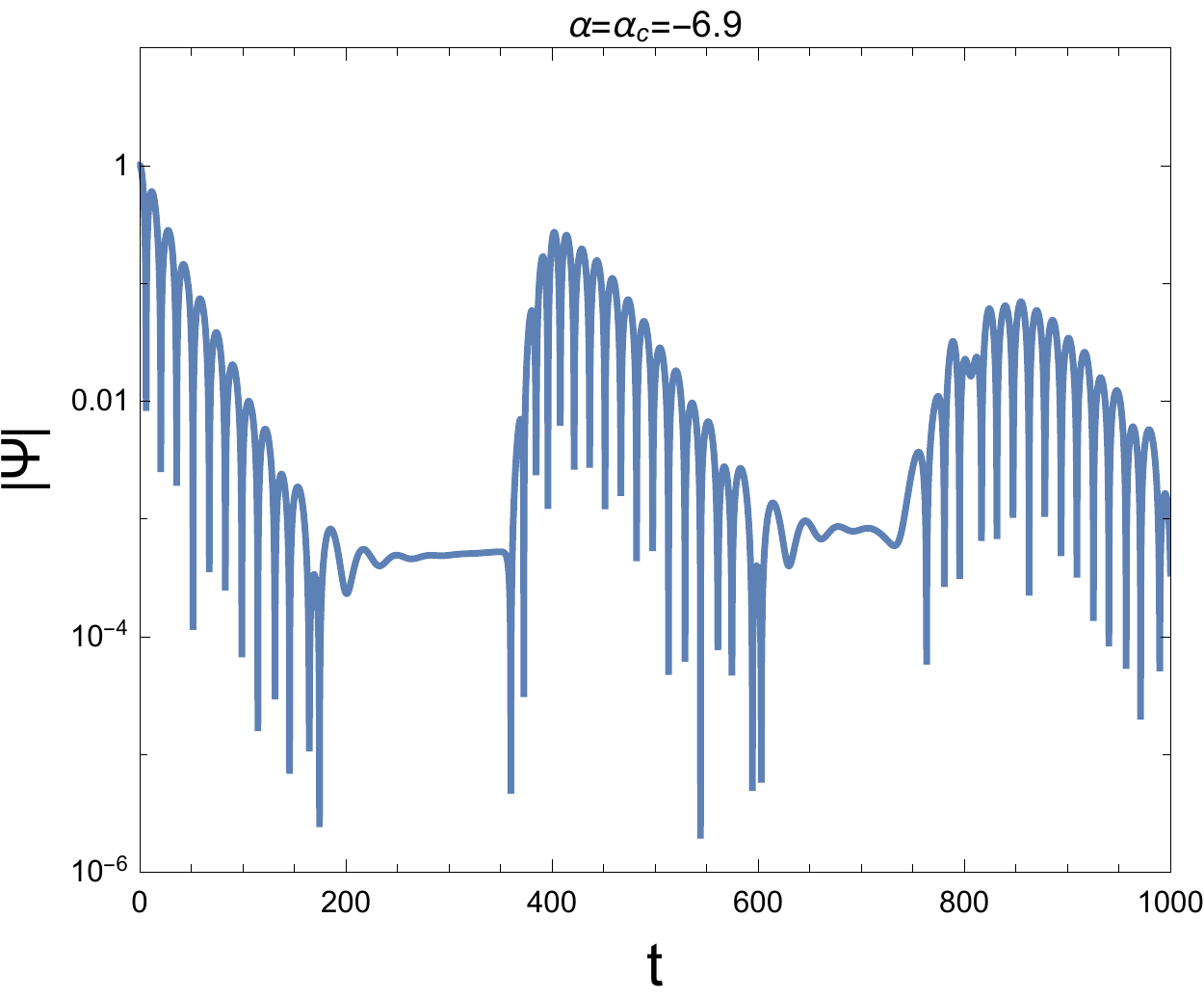}\quad
	\includegraphics[width=0.45\textwidth]{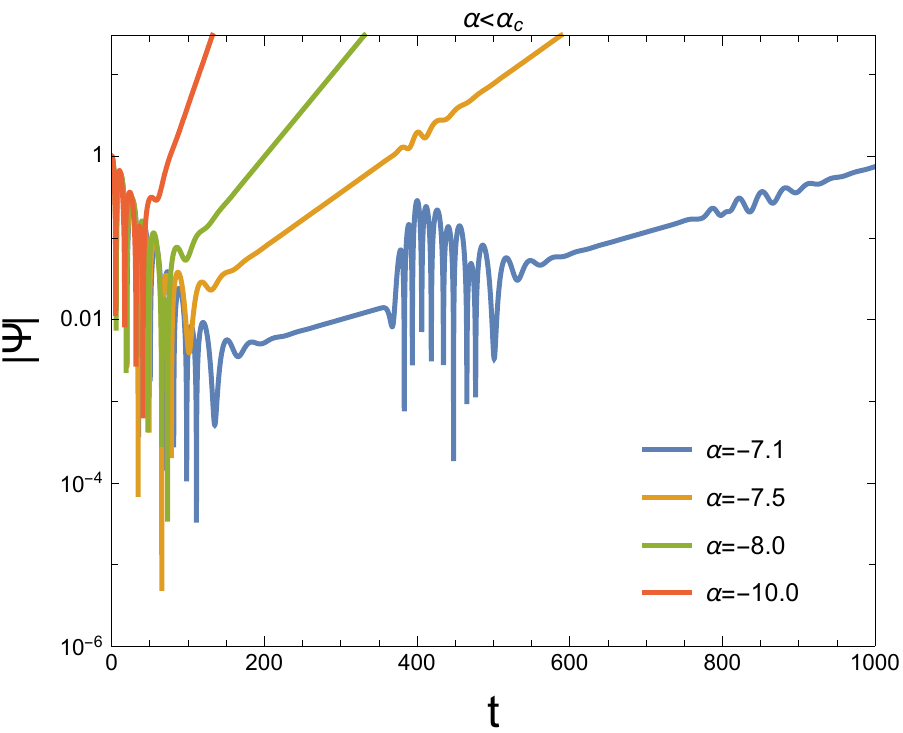}
	\caption{(color online) Time evolutions of the scalar field perturbation for  $Q=0.8, B=0.1$ and various values of $\alpha$. From left to right and top to bottom, $|\alpha|$ increases from $0$ to $10.0$.} \label{fig:AlphaNegative}
\end{figure}

Let us first consider the $\alpha<0$ case. Without loss of generality, we fix $Q=0.8$ and $B=0.1$ \footnote{Typical value of the magnetic field we considered in this work is $B M \sim 0.1$. In our units $c=G =4\pi \epsilon_0 =1$, we have the following relation
\begin{equation}
	\frac{1}{M} \simeq 2.36 \times 10^{19} \left(\frac{M_\odot}{M}\right) \textrm{Gauss}, \nonumber
\end{equation}
where $M_\odot$ is the solar mass. So $B M \sim 0.1$ corresponds to a magnetic field $B \sim 2.36 \times 10^{18} \left(\frac{M_\odot}{M}\right) \textrm{Gauss}$; For stellar-mass BHs with $M \sim 10 M_\odot$, $B \sim 10^{17}\textrm{Gauss}$, while for supermassive BHs with $M \sim 10^6 M_\odot$ (for example the Sagittarius A$^\ast$), $B \sim 10^{12} \textrm{Gauss}$. In general, for BHs with mass $M >10^2 M_\odot$, the magnetic field considered will be smaller than the ever-measured strongest magnetic field $B \sim 10^{16} \textrm{Gauss}$ \cite{Olausen:2013bpa}. For smaller magnetic field, the tachyonic instability and spontaneous scalarization can still be triggered as long as the coupling constant $\alpha$ is large enough.}. In Fig. \ref{fig:AlphaNegative}, time evolutions of the scalar field perturbations for various values of the coupling constant $\alpha$ are shown. For the figure, one can see that there exists a threshold value of the coupling constant $|\alpha| = |\alpha_c| \simeq 6.9$ above which the tachyonic instability is triggered which is a signal of spontaneous scalarization. With the increase of $|\alpha|$, the instability appears earlier and becomes more violent. This can be understood as larger $|\alpha|$ will yield a more negative $m_{\rm eff}^2$. 

Moreover, when $|\alpha| < |\alpha_c|$, from the waveforms one can observe an interesting phenomenon: two different types of ringdown modes appear at different stages. The first ringdown modes, appearing at early times, are being in fact similar to the quasinormal modes (QNMs) of RN BH under a massive scalar field perturbation \cite{Galtsov:1978ag,Konoplya:2007yy,Konoplya:2008hj}. After time of order $t \sim \frac{1}{B}$, the ``Melvin-like" modes with smaller and smaller amplitudes are excited as a result of an effective "wall" induced by the magnetic field \cite{Brito:2014nja}. This interesting phenomenon has already been observed in GR \cite{Brito:2014nja} and dCSG theory \cite{Zhang:2022sgt}. Similar phenomenon was also observed in the context of "dirty" BHs \cite{Barausse:2014tra}. When $|\alpha| > |\alpha_c|$, unstable tachyonic modes are excited which will occupy the energy of the system. Then the Melvin-like modes are suppressed and even disappear when $|\alpha|$ is large enough.

\begin{figure}[!htbp]
	\centering
	\includegraphics[width=0.45\textwidth]{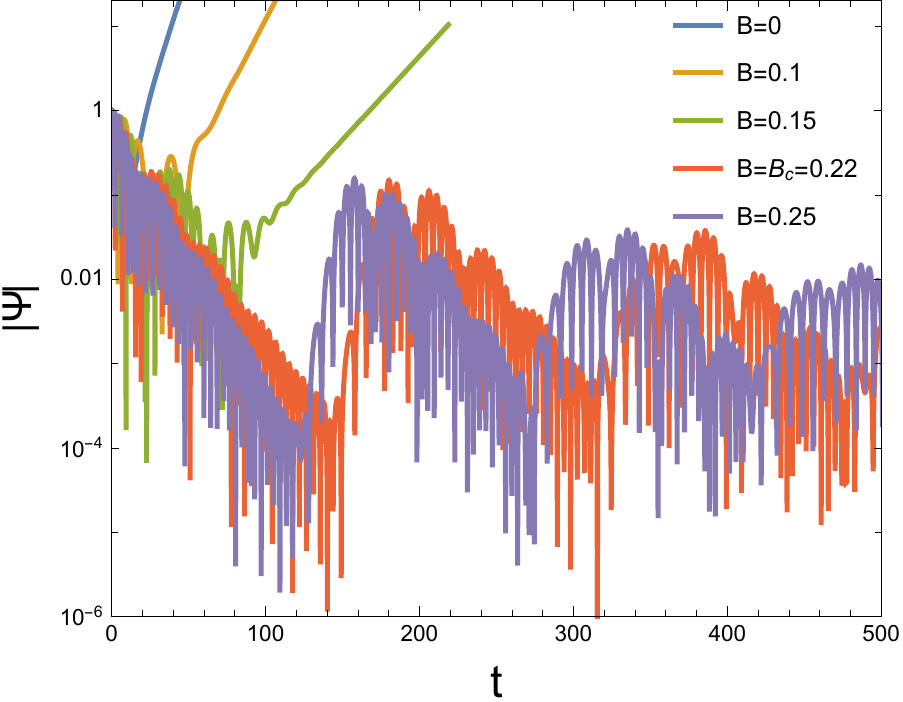}\quad
	\includegraphics[width=0.45\textwidth]{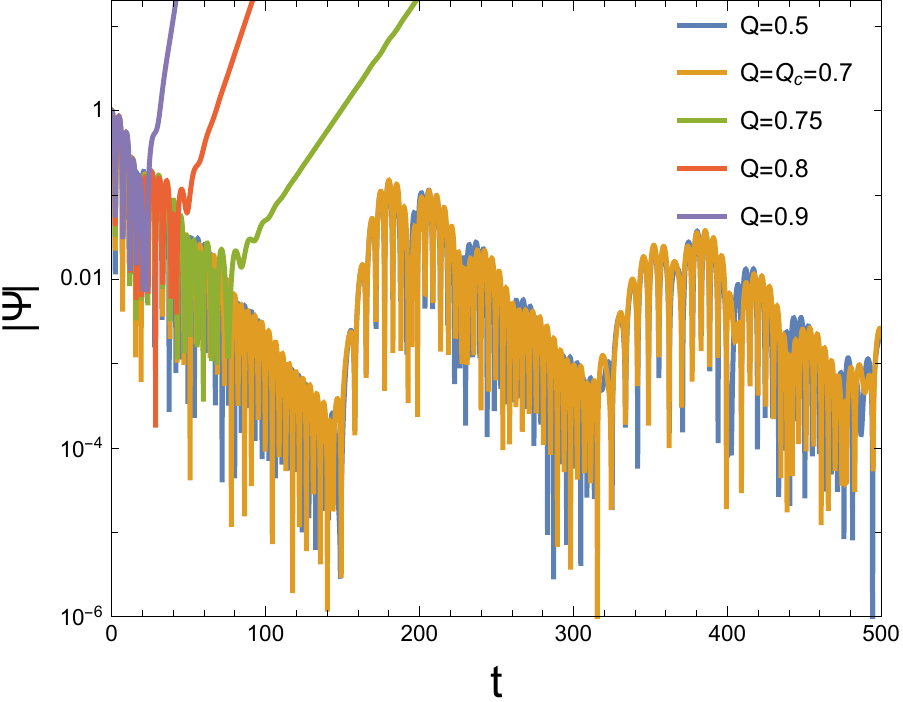}
	\caption{(color online) Time evolutions of the scalar field perturbation for {\em Left}: $Q=0.7, \alpha=-20$ and various values of $B$ and {\em Right}: $B=0.22, \alpha=-20$ and various values of $Q$.}
	\label{fig:BQEffectAlphaNegative}
\end{figure}

\begin{figure}[!htbp]
	\centering
	\includegraphics[width=0.45\textwidth]{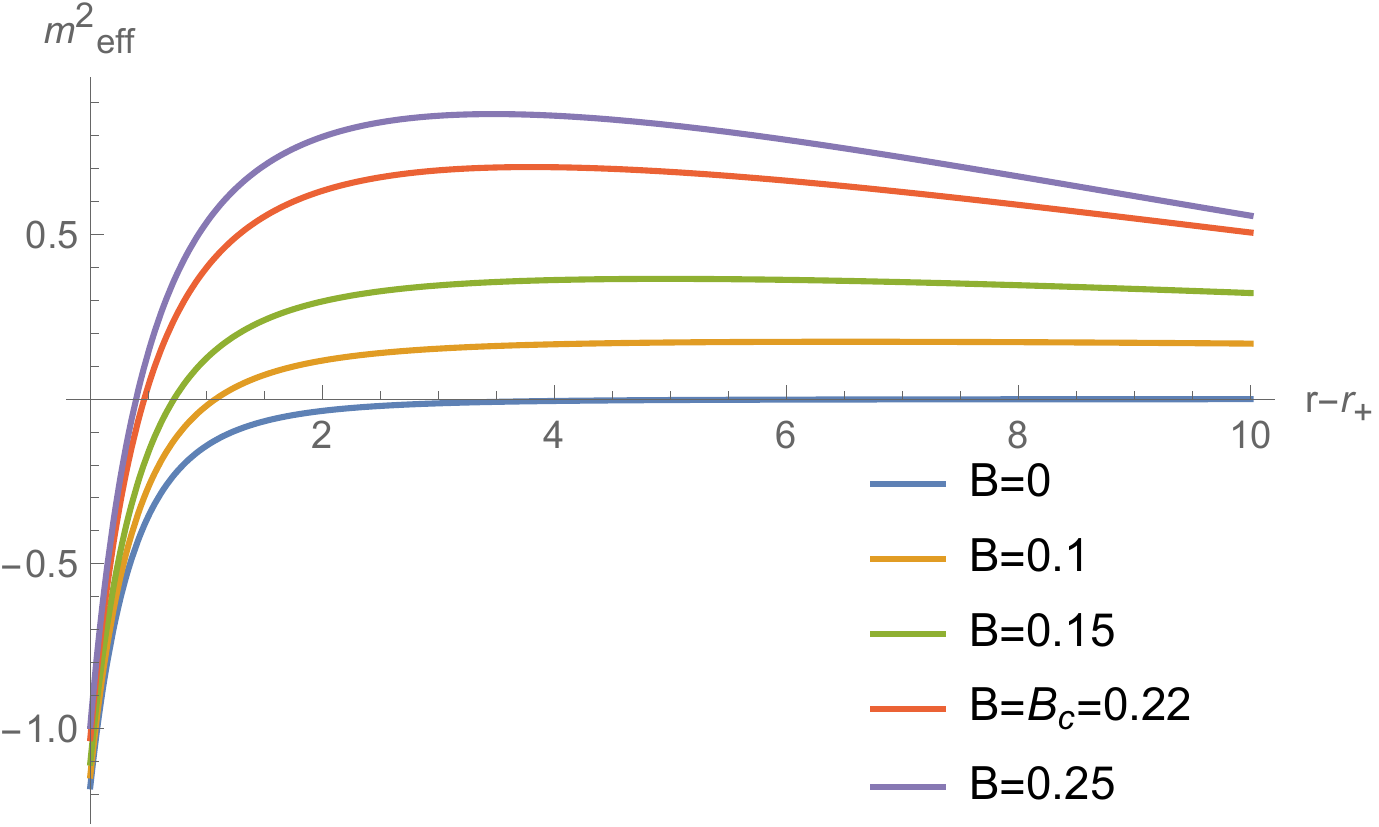}\quad
	\includegraphics[width=0.45\textwidth]{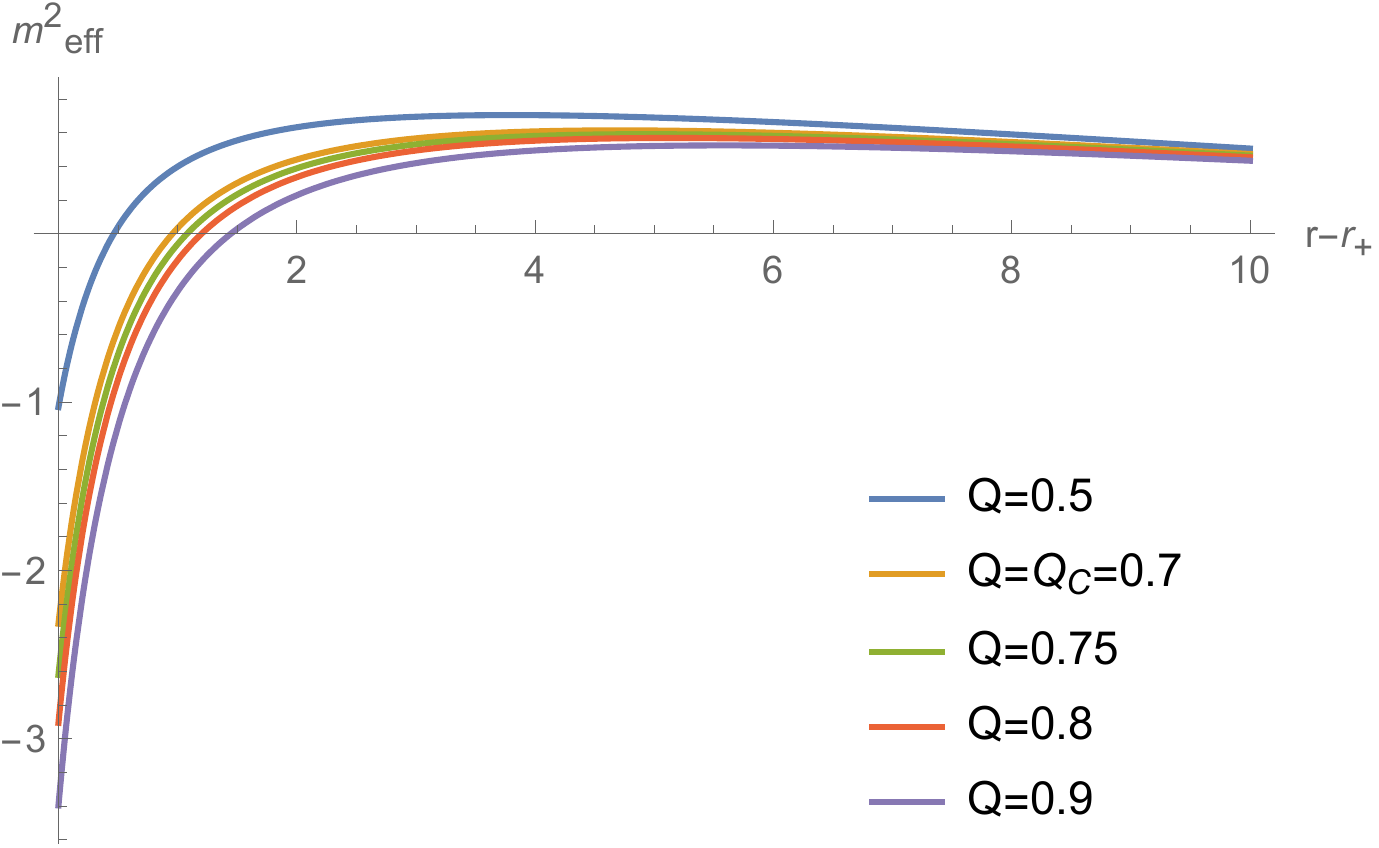}
	\caption{(color online) Profiles of the effective mass square $m_{\textrm{eff}}^2 = -\frac{1}{2} \alpha F^2$ for fixed $\theta=\frac{\pi}{10}$. {\em Left}: $Q=0.7$ while $B$ is varied. {\em Right}: $B=0.22$ while $Q$ is varied.}
	\label{fig:MassSquareAlphaNegative}
\end{figure}

To study the influences of the magnetic field $B$ and BH charge $Q$ on the instability, we fix $\alpha$ and plot the time evolutions of the scalar field perturbations for various values of $B$ and $Q$ in Fig. \ref{fig:BQEffectAlphaNegative}. From the left panel, one can see that there exists a threshold $B = B_c \simeq 0.22$ below which the tachyonic instability is triggered. And the tachyonic instability appears later and becomes more mild as $B$ is increased. This means that the presence of the magnetic field suppresses the tachyonic instability and even quenches it when $B$ is large enough. This can be understood from the  left panel of Fig. \ref{fig:MassSquareAlphaNegative}, from which one can see that larger $B$ makes the $m_{\rm eff}^2$ less negative near the horizon. Moreover, for $B>B_c$, the Melvin-like modes appears earlier for larger $B$, which can be understood as the effective "wall" becomes closer to the horizon. 

From the right panel of Fig. \ref{fig:BQEffectAlphaNegative}, one can see that the influence of the BH charge $Q$ on the tachyonic instability is opposite to that of $B$: larger $Q$ makes the instability appears earlier and becomes more violent. Also, this can be understood from the right panel of Fig. \ref{fig:MassSquareAlphaNegative}, from which one can see that larger $Q$ makes the $m_{\rm eff}^2$ more negative near the horizon.

\subsection{$\alpha>0$}

\begin{figure}[!htbp]
	\includegraphics[width=0.7\textwidth]{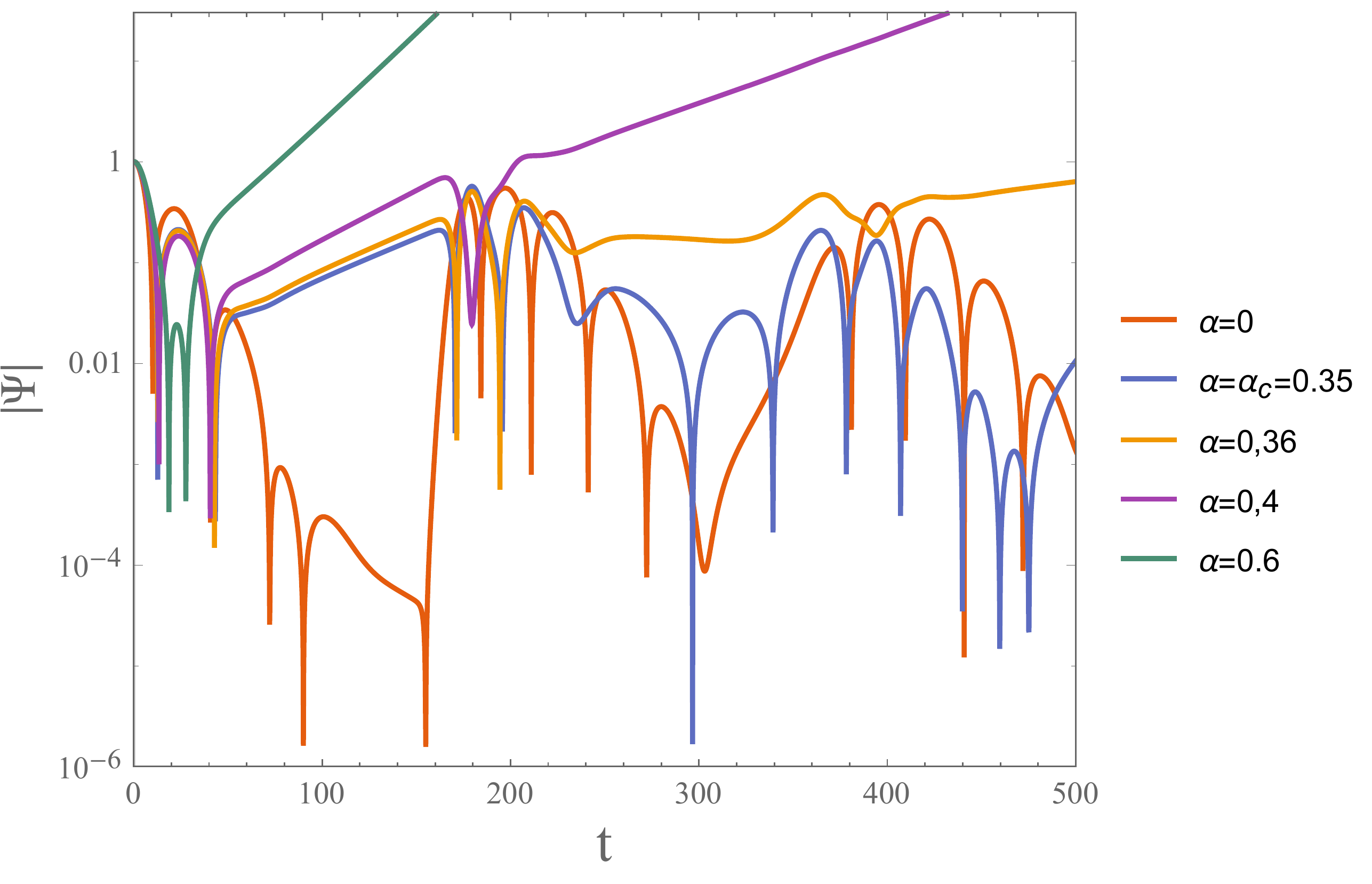}
	\caption{(color online) Time evolutions of the scalar field perturbation for $Q=0.2, B=0.2$ and various values of $\alpha$. From left to right and top to bottom, $\alpha$ increases from $0$ to $0.6$.}\label{fig:AlphaPostive}
\end{figure}

\begin{figure}[!htbp]
	\centering
	\includegraphics[width=0.45\textwidth]{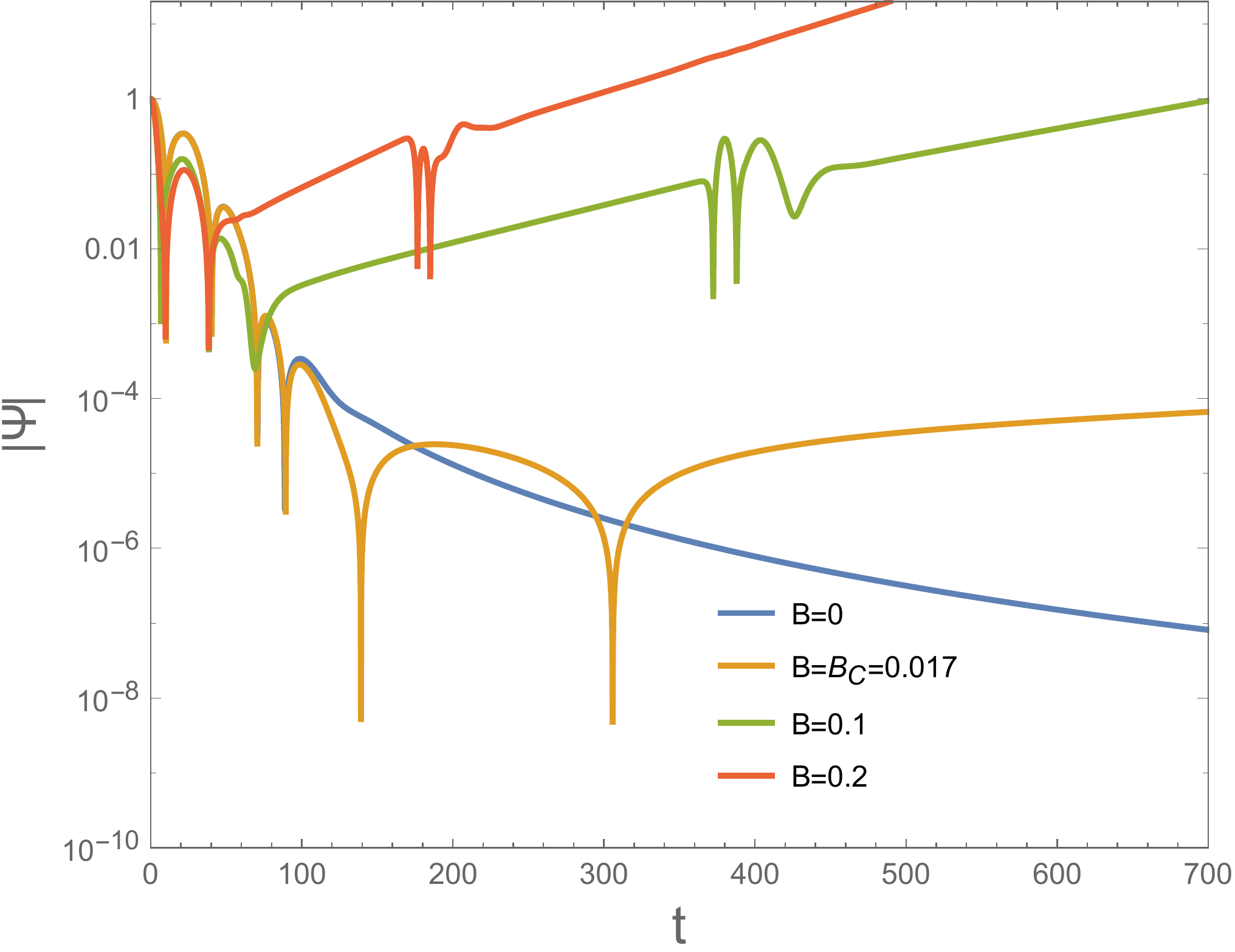}\quad
	\includegraphics[width=0.45\textwidth]{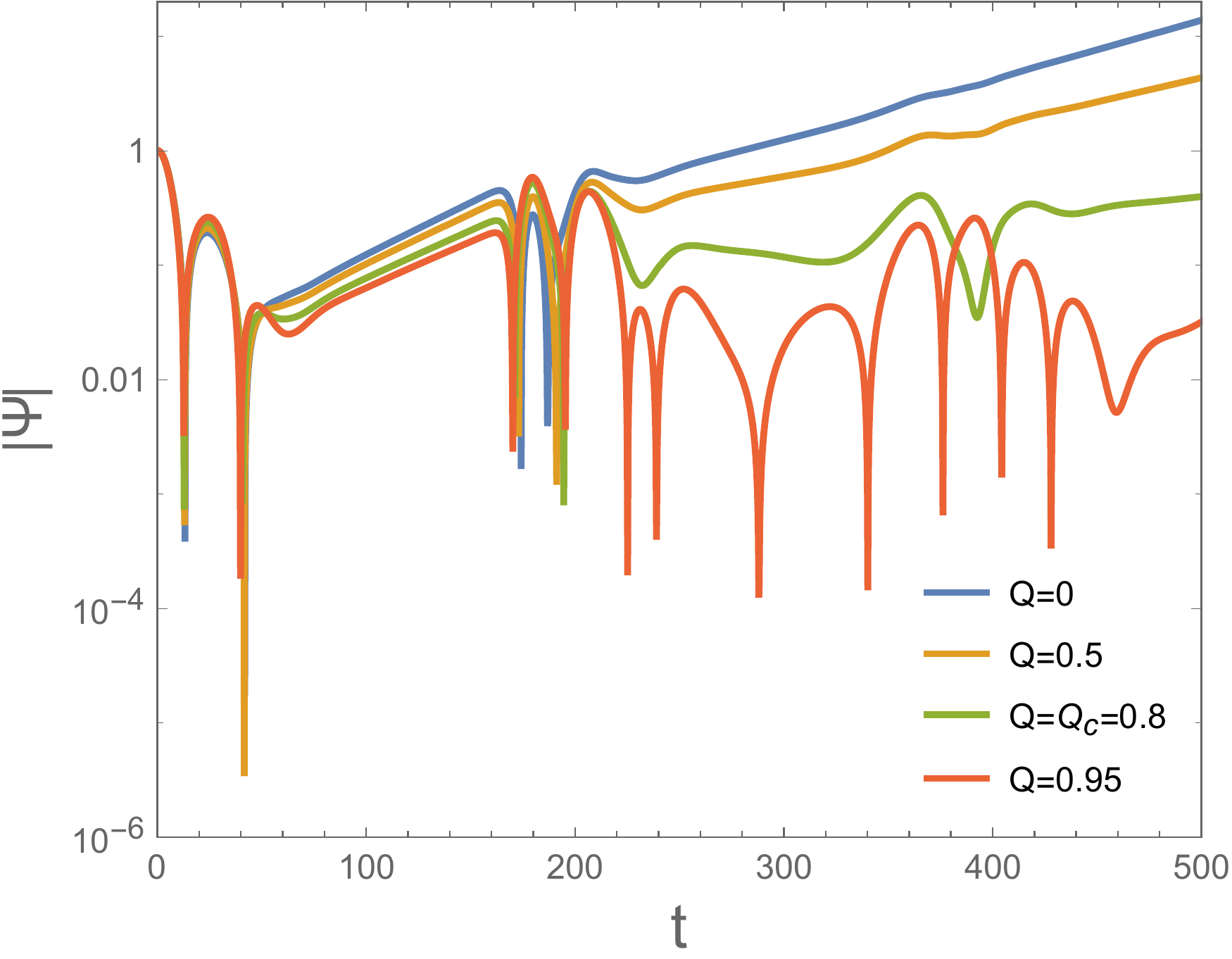}
	\caption{(color online) Time evolutions of the scalar field perturbation for {\em Left}: $Q=0.4, \alpha=0.4$ and various values of $B$ and {\em Right}: $B=0.2, \alpha=0.38$ and various values of $Q$.}
	\label{fig:BQEffectAlphaPositive}
\end{figure}

\begin{figure}[!htbp]
	\centering
	\includegraphics[width=0.45\textwidth]{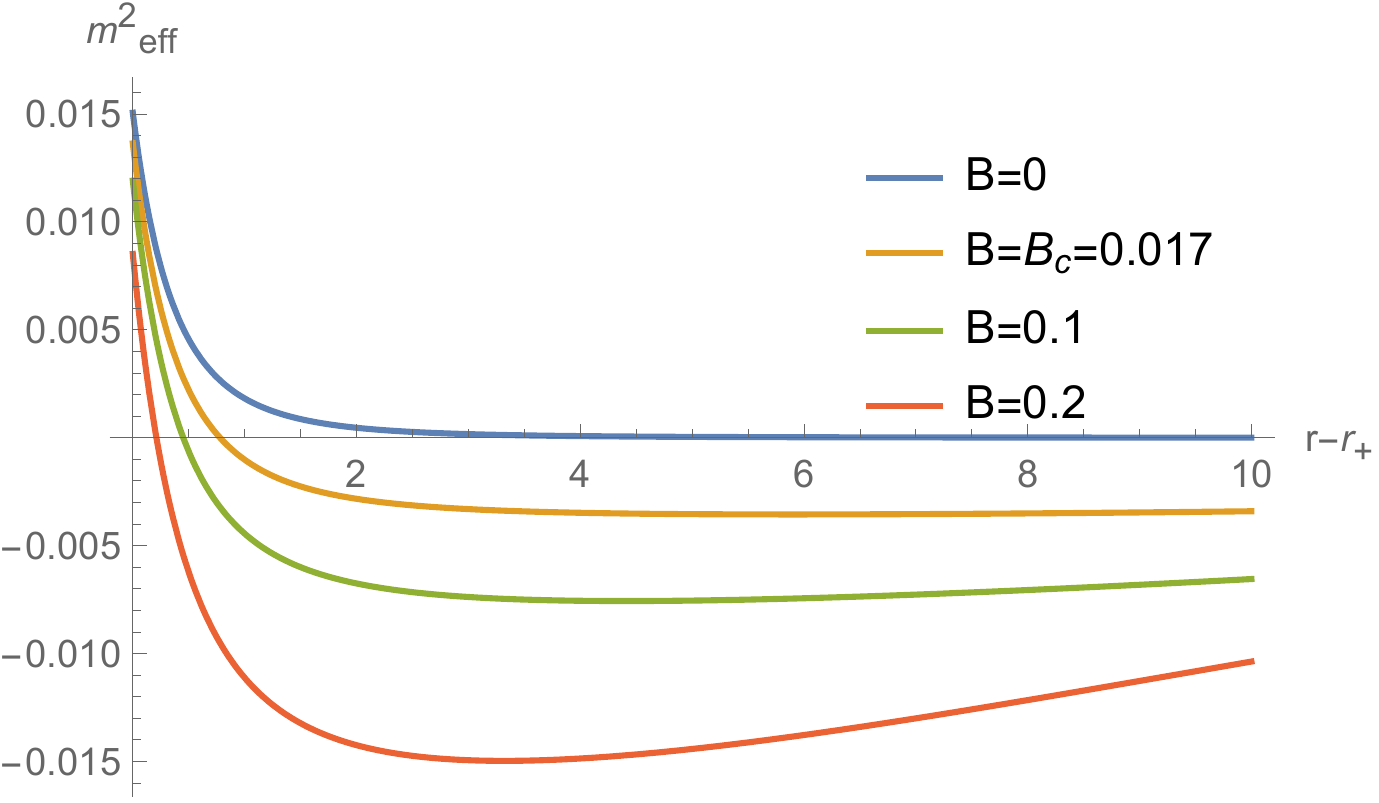}\quad
	\includegraphics[width=0.45\textwidth]{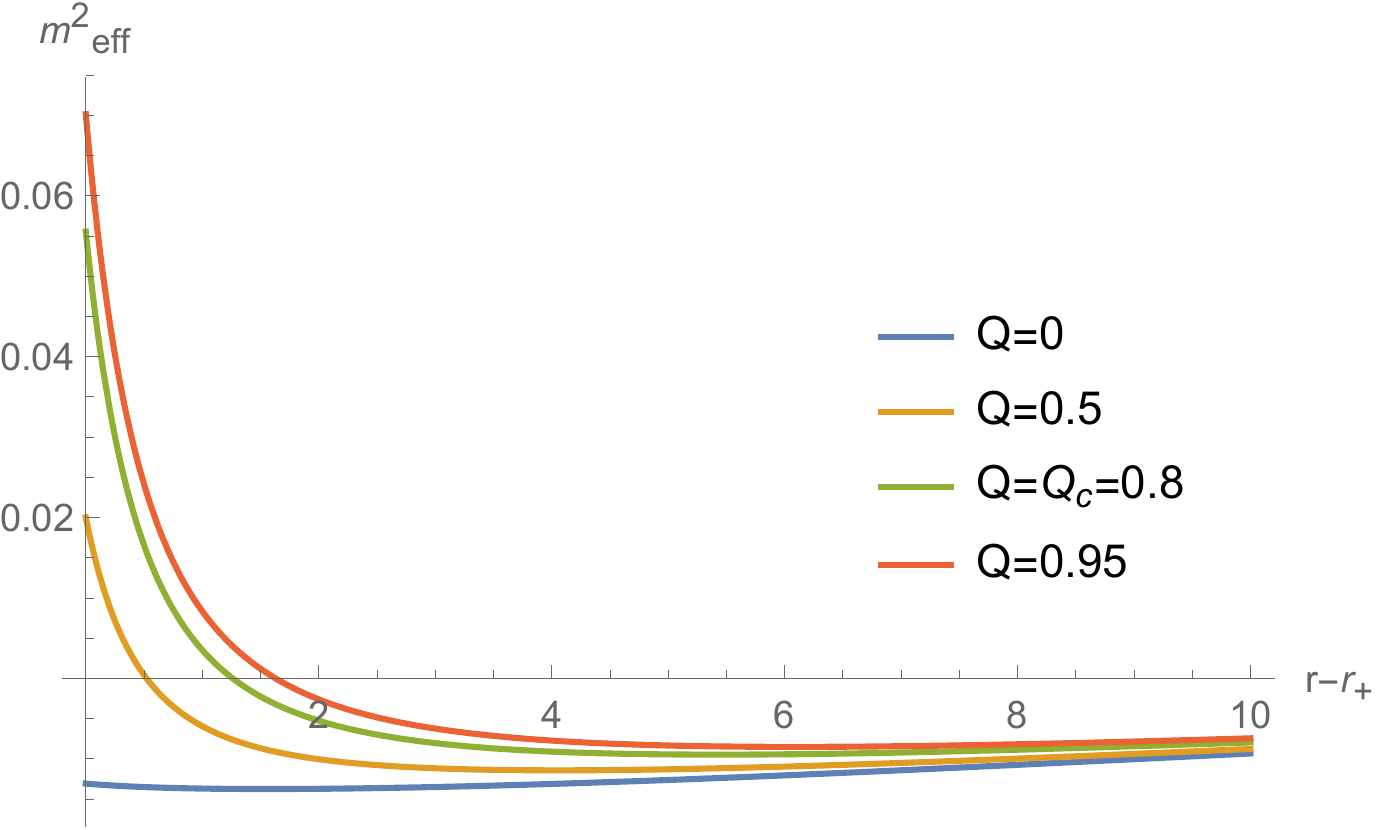}
	\caption{(color online) Profiles of the effective mass square $m_{\textrm{eff}}^2 = -\frac{1}{2} \alpha F^2$ for fixed $\theta=\frac{\pi}{10}$. {\em Left}: $Q=0.4$ while $B$ is varied. {\em Right}: $B=0.2$ while $Q$ is varied.}
	\label{fig:MassSquareAlphaPositive}
\end{figure}

Now we turn to the $\alpha>0$ case. In Fig. \ref{fig:AlphaPostive}, time evolutions of the scalar field perturbations are shown for fixed $Q=0.2, B=0.2$ and various values of $\alpha$. As the $\alpha<0$ case, the tachyonic instability will occur when $\alpha$ exceeds a threshold value. The influences of $B$ and $Q$ on the wave dynamics are shown in Fig. \ref{fig:BQEffectAlphaPositive}, from which one see that larger $B$ or smaller $Q$ makes the instability appears earlier and becomes more violent. This is different from the $\alpha<0$ case, and can also be understood qualitatively from the behaviors of $m_{\rm eff}^2$. From Fig. \ref{fig:MassSquareAlphaPositive}, one can see that $m_{\rm eff}^2$ is positive near the horizon but becomes negative away from the horizon. Larger $B$ or smaller $Q$ makes $m_{\rm eff}^2$ more negative indicating more violent instability. This also implies that  the unstable tachyonic modes are excited away from the horizon rather than near the horizon.

\section{Summary and Discussions}

In this work, we study the scalar field perturbations of the RNM BHs in the framework of EMs theory by performing time simulations. In this framework, the tacyonic instability and associated spontaneous scalarization are induced by the matter field rather than the curvature. Our focus is on the influences of the magnetic field on the wave dynamics. In the absence of the magnetic field, the tachyonic instability can only be triggered for negative coupling constant $\alpha$. However, while the magnetic field is present, it can occur for either sign of $\alpha$. Moreover, for different sign of $\alpha$, the influences of the magnetic field on the tachyonic instability will be rather different. For $\alpha<0$, the effective mass square $m_{\textrm{eff}}^2$ becomes negative near the horizon and the unstable tachyonic modes should be excited there, and larger $B$ will make the instability appears later and more mild or even disappears when $B$ is large enough; While for $\alpha>0$, the effective mass square $m_{\textrm{eff}}^2$ is positive near the horizon but becomes negative away from the horizon thus the unstable modes are instead excited away from the horizon, and the influences of the magnetic field are opposite to that in $\alpha<0$ case. Comparing to the studies in EsGB \cite{Annulli:2022ivr} and dSCG \cite{Zhang:2022sgt}, one can have the conclusion that the influences of the magnetic field are model-dependent. 

As the presence of the magnetic field changes the asymptotic structures of the spacetime, the waveforms are shown to be greatly different from that of $B=0$. Beyond the usual QNMs modes appearing in early times, a new type of ringdown modes--the ``Melvin-like" modes--are excited at late times. The influences of the BH charge $Q$ on the instability are also studied in detail, and also differs for different sign of $\alpha$. 

In this work, we only perform the time simulations of the wave dynamics at linear level. To get a complete picture of the process of spontaneous scalarization, a full non-linear time evolution of the system including the spacetime is called for. Also, it is interesting and important to obtain the final scalarized BH metric and study its various properties, such as its astrophysical implications. We would like to leave these questions for further investigations.

\begin{acknowledgments}

	This work is supported by the National Natural Science Foundation of China (NNSFC) under Grant No. 12075207.
\end{acknowledgments}

\end{document}